\documentclass[aps,twocolumn,showpacs,preprintnumbers,superscriptaddress,amsmath,amssymb]{revtex4}

\usepackage{graphicx}
\usepackage{dcolumn}
\usepackage{bm}

\begin{document}

\title{Doping Effects on the two-dimensional Spin Dimer Compound SrCu$_2$(BO$_3$)$_2$}

\author{G. T. liu}
\author{J. L. Luo}
\email{JLLuo@aphy.iphy.ac.cn}
\affiliation{Institute of Physics,
Chinese Academy of Sciences, P.O. Box 603, Beijing 100080, P. R.
China}

\author{T. Xiang}
\affiliation{Institute of Theoretical Physics and
Interdisciplinary Center of Theoretical Studies, Chinese Academy
of Science, P.O. Box 2735, Beijing 100080, P. R.  China}

\author{N. L. Wang}
\affiliation{Institute of Physics, Chinese Academy of Sciences,
P.O. Box 603, Beijing 100080, P. R. China}

\author{Z. H. Wu}
\affiliation{Department of Physics, Chongqing University,
Chongqing 400044, P. R. China}

\author{X. N. Jing}
\affiliation{Institute of Physics, Chinese Academy of Sciences,
P.O. Box 603, Beijing 100080, P. R. China}

\author{D. Jin}
\affiliation{Institute of Physics, Chinese Academy of Sciences,
P.O. Box 603, Beijing 100080, P. R. China}

\date{\today}

\begin{abstract}
A series of compounds M$_{0.1}$Sr$_{0.9}$Cu$_2$(BO$_3$)$_2$ with
Sr substituted by M=Al, La, Na and Y were prepared by solid state
reaction. XRD analysis showed that these doping compounds are
isostructural to SrCu$_2$(BO$_3$)$_2$. The magnetic susceptibility
from 1.9K to 300K in an applied magnetic field of 1.0T and the
specific heat from 1.9K to 25K in applied fields up to 14T were
measured. The spin gap is deduced from the low temperature
susceptibility as well as the specific heat. It is found that the
spin gap is strongly suppressed by magnetic fields. No
superconductivity is observed in all four samples.
\end{abstract}

\pacs{Valid PACS appear here}
\maketitle

\section{\label{sec:level1}INTRODUCTION}

During the past few years, a number of novel quantum spin systems
have been discovered experimentally. Among them, the spin ladder
compounds, such as SrCu$_2$O$_3$ \cite{Azuma} and CaV$_2$O$_5$
\cite{Iwase}, and the dimerised spin gap compound
SrCu$_2$(BO$_3$)$_2$ found by Kageyama $et$ $al$. \cite{Kageyama1}
in two dimensions have attracted great attention recently.
SrCu$_2$(BO$_3$)$_2$ was first synthesized by Smith $et$ $al$.
\cite{Smith}. It has a tetragonal unit cell with lattice
parameters $a=b=8.995$\AA\ and $c=6.649$\AA\ at room temperature.
In this compound, Sr and CuBO$_3$ planes are stacked along the c
axis alternatively. In the CuBO$_3$ plane, a neighboring pair of
rectangular planar CuO$_4$ forms a spin dimer and the dimers are
interconnected by triangular planar BO$_3$. The two-dimensional
coordinates of the Cu$^{2+}$ spins were shown in Fig. 1.

\begin{figure}
\includegraphics[width=7cm]{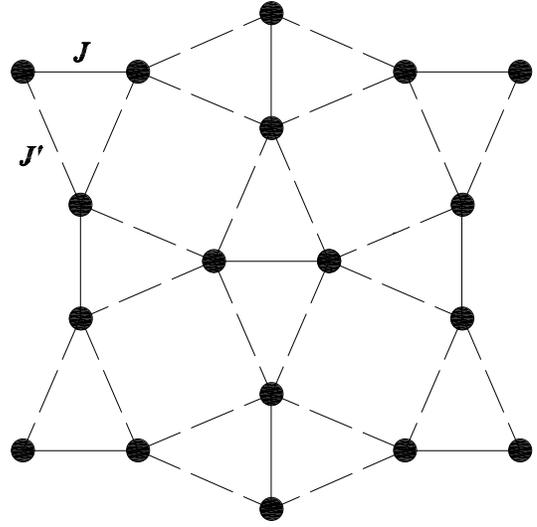}
\caption{\label{fig:liufig_1} {Lattice structure of the Cu$^{2+}$
spins of SrCu$_2$(BO$_3$)$_2$. The nearest-neighbor bonds are
represented by solid lines and the next-nearest-neighbor ones by
dashed lines.}}
\end{figure}

Miyahara $et$ $al.$ \cite{Miyahara} found that the in-plane
coupling of spins in this system is topologically equivalent to
that in the Shastry-Sutherland model \cite{Shastry1} and the
magnetic properties of SrCu$_2$(BO$_3$)$_2$ can be well described
by this model. The Shastry-Sutherland model is a special case of
the frustrated Heisenberg model defined by the Hamiltonian
\begin{equation}
\emph{H}=\emph{J}\sum_{nn}\emph{S}_i\emph{S}_j +
\emph{J$^{'}$}\sum_{nnn}\emph{S}_i\emph{S}_j
 \end{equation}
where nn (nnn) stands for the nearest (next-nearest) neighbor
spins, and $J$ ($J^{'}$) stands for the next-nearest inter-dimer
(intra-dimer) coupling. The Shastry-Sutherland model corresponds
the case $J/J^\prime<0.68$ \cite{Miyahara, Shastry1}, whose ground
state is a dimerised spin singlet. In the limit
$J/J^\prime\rightarrow 0$ or $J/J^\prime \rightarrow \infty$, the
ground state of the frustrated Heisenberg model is
antiferromagnetically long range ordered and low-lying spin
excitations are critical. In the intermediate $J/J^\prime$ regime,
the ground state becomes non-critical and the spin excitations are
gapped. With decreasing $J/J^\prime$, a phase transition from a
gapless to a gapped spin singlet state is shown to take place at a
critical value $(J/J^{'})_c \sim 0.7$ \cite{Miyahara, Zheng}. The
ratio of $J/J^{'}$ in SrCu$_2$(BO$_3$)$_2$ is about 0.68
\cite{Miyahara}, a little bit smaller than the critical value.
This suggests that the ground state of SrCu$_2$(BO$_3$)$_2$ is in
a spin-gapped state.

Recently, Shastry and Kumar studied the doped Shastry-Sutherland
system from the RVB theory of the t-J-like model. They found that
the Mott-Hubbard gap will collapse upon doping, and the
pre-existing spin pairs will propagate freely, leading to
superconductivity \cite{Shastry2}. Similar conclusion was reached
by Kimura and co-workers. However, they found that
superconductivity is more favored around the quarter filling
rather than the half filling \cite{Kimura}.

In this paper, we report the magnetic susceptibility and specific
heat measurements on the dimer compounds SrCu$_2$(BO$_3$)$_2$ and
M$_{0.1}$Sr$_{0.9}$Cu$_2$(BO$_3$)$_2$ (M=Al, La, Na and Y). The
spin excitation gaps of M$_{0.1}$Sr$_{0.9}$Cu$_2$(BO$_3$)$_2$ are
deduced from the low temperature data and found to be close to
that of SrCu$_2$(BO$_3$)$_2$. The gap is strongly suppressed by
the applied fields. However, no superconductivity is observed in
these systems.

\section{\label{sec:level1}Synthesis and Experimental Details}

The polycrystalline materials SrCu$_2$(BO$_3$)$_2$ and
M$_{0.1}$Sr$_{0.9}$Cu$_2$(BO$_3$)$_2$ with M=Al, La, Na and Y were
synthesized by solid state reaction. Stoichiometric amounts of
high-purity SrCO$_3$, CuO, B$_2$O$_3$, Y$_2$O$_3$, La$_2$O$_3$,
NaO and Al$_2$O$_3$ powders were ground and mixed, then pressed
with a pressure of ~10MPa into pellets and heated in air at 830
$^{\circ}$C for 4 hours. The resulting pellets were then reground,
repelletized and sintered at 870 $^{\circ}$C  in oxygen atmosphere
for one week.

\begin{figure}
\includegraphics[width=7cm]{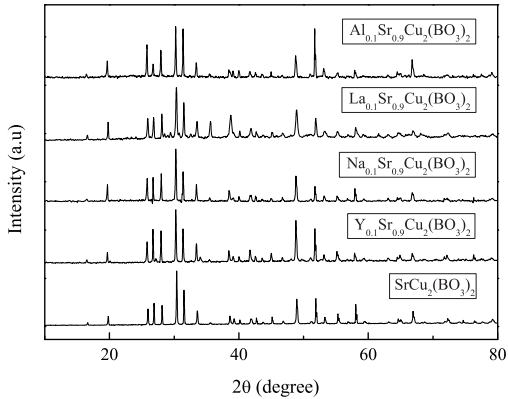}
\caption{\label{fig:liufig_2} {XRD patterns for
 M$_{0.1}$Sr$_{0.9}$Cu$_2$(BO$_3$)$_2$ (M=Al, La, Na and Y) and SrCu$_2$(BO$_3$)$_2$
 polycrystalline compounds.}}
\end{figure}

X-ray diffraction measurements (XRD) were carried out on powder
samples of SrCu$_2$(BO$_3$)$_2$ and
M$_{0.1}$Sr$_{0.9}$Cu$_2$(BO$_3$)$_2$. Fig. 2 shows the XRD
patterns for all samples. The patterns indicate that they are all
single-phase products. From the precise position of $2\theta$, the
lattice parameters were obtained and shown in Table 1. The
in-plane parameters $a$ and $b$ of all the four samples
M$_{0.1}$Sr$_{0.9}$Cu$_2$(BO$_3$)$_2$ and $c$ of La and Na
substituted samples are almost unchanged in comparison with
SrCu$_2$(BO$_3$)$_2$. However, the $c$-axis lattice parameters $c$
of Al and Y substituted samples are changed notably. This is
because the ionic radius of La$^{3+}$ (1.06\AA) and Na$^+$
(0.97\AA) are very close to that of Sr$^{2+}$ (1.12\AA), while
Al$^{3+}$ (0.51\AA) and Y$^{3+}$ (0.89\AA) are considerably
smaller than it. These results indicate that the doped elements
are located between CuBO$_3$ layers. It leads to the variation of
interlayer spacing and leaves the CuBO$_3$ plane almost unchanged.

\begin{figure}[b]
\includegraphics[width=7cm]{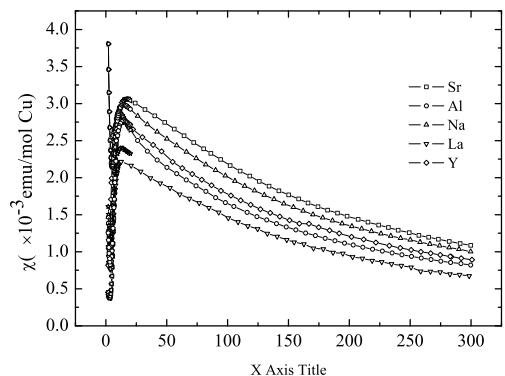}
\caption{\label{fig:liufig_3} {Temperature dependence of the
magnetic susceptibility in M$_{0.1}$Sr$_{0.9}$Cu$_2$(BO$_3$)$_2$
(M=Al, La, Na, Y) and SrCu$_2$(BO$_3$)$_2$ polycrystalline
compounds in a field of 1.0T.}}
\end{figure}

The magnetic susceptibility and the specific heat measurements
were performed in commercial Quantum Design PPMS. The
susceptibility was measured using a vibrating sample magnetometer.
The specific heat was measured using the relaxation method. The
field dependence of thermometer and addenda was carefully
calibrated before the specific heat was measured.

All doped materials M$_{0.1}$Sr$_{0.9}$Cu$_2$(BO$_3$)$_2$ are
insulators. Their resistances are all above 200M$\Omega$ at room
temperature.

\begin{figure}[b]
\includegraphics[width=7cm]{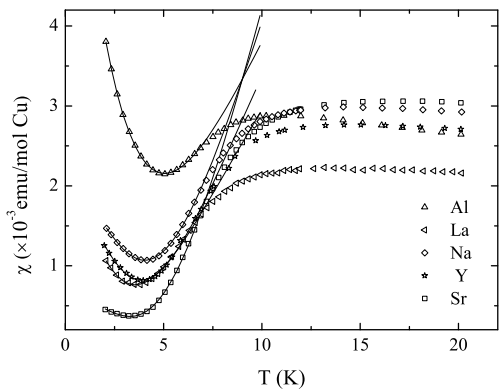}
\caption{\label{fig:liufig_4}{Susceptibility $\chi$ versus
temperature $T$ below 20K for
M$_{0.1}$Sr$_{0.9}$Cu$_2$(BO$_3$)$_2$ (M=Al, La, Na, Y) and
SrCu$_2$(BO$_3$)$_2$ compounds. The solid lines are fitting curves
with Eq. (\ref{chi}).}}
\end{figure}

\section{\label{sec:level1}Results and Discussions\protect\\}

Fig. 3 shows the magnetic susceptibility $\chi$=$M/H$ as a
function of temperature $T$ for SrCu$_2$(BO$_3$)$_2$ and
M$_{0.1}$Sr$_{0.9}$Cu$_2$(BO$_3$)$_2$ in an applied field of 1.0T
from 1.9K to 300K. For all samples, $\chi$ follows well the
Curie-Weiss law in high temperatures. A broad peak appears around
15K. In low temperatures, a spin gap opens and $\chi$ drops
sharply below 10K (Fig. 4). Below 4K, a small Curie-Weiss-like
upturn is observed. This upturn is the contribution of magnetic
impurities or defects \cite{Kageyama1}.

The excitation gap $\Delta$ can be determined from the temperature
dependence of $\chi$ in low temperatures. In general, the
low-lying spin excitation spectrum around the gap minima has
approximately the form \cite{Tao,Sorensen}:
\begin{equation}
{\varepsilon(k)} \sim \Delta+\alpha \left({\bf k}-{\bf
Q}_0\right)^2,
\end{equation}
where ${\bf Q}_0$ is the vector where the minimum gap is located.
In the limit $T\ll\Delta$, it can be readily shown that the spin
susceptibility and the specific heat are respectively given by
\begin{eqnarray}
{\chi} & \sim  & \exp(-\Delta/T), \\
 C & \sim  & \frac{1}{T}\exp(-\Delta/T).
\end{eqnarray}
Adding the contribution from magnetic impurities or defects and
the contribution from ion cores, the total susceptibility in low
temperatures can be expressed as
\begin{equation}
{\chi}=\frac{C^\prime}{T-\theta^\prime}+a\exp(-\Delta/T)+\chi_{0}.
\label{chi}
\end{equation}
The first term on the right hand side is the contribution of
magnetic impurities and $\theta^\prime$ is the Curie temperature.
The last term is the diamagnetic contribution from ion cores.

The low temperature susceptibility data can indeed be described by
Eq. (\ref{chi}). Our fitting curves are shown by solid lines in
Fig. 4. The values of $\chi_0$, $ C^{'}$ and $\Delta$ are listed
in Table 1. The spin gap is 21.5K for SrCu$_2$(BO$_3$)$_2$,
consistent with the result of Kageyama \cite{Kageyama1}. By
comparison with SrCu$_2$(BO$_3$)$_2$, we find that the spin gap is
suppressed by doping. The suppression of the energy gap is most
apparent for M = Y and La samples. The spin gap in the La doped
material is only 14.1K.

\begin{figure}[b]
\includegraphics[width=7cm]{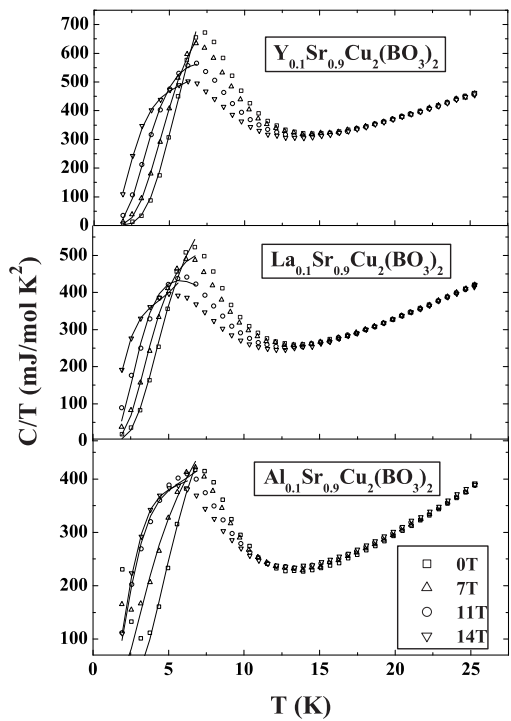}
\caption{\label{fig:liufig_5} {$C/T$ for M$_{0.1}$Sr$_{0.9}$Cu$_2$(BO$_3$)$_2$ (M= Y,
La and Al) polycrystalline compounds measured at $H$=0, 7, 11 and
14T.}}
\end{figure}

\renewcommand\arraystretch{1.5}
\begin{table*}
\caption{\label{tab:Table1}Values of lattice parameters $(a, b,
c)$, the diamagnetic susceptibility of ion cores $\chi_0$, $C^{'}$
and $\Delta$ for SrCu$_2$(BO$_3$)$_2$ and
M$_{0.1}$Sr$_{0.9}$Cu$_2$(BO$_3$)$_2$. $\Delta^{1}$ and
$\Delta^{2}$ represent the spin gap deduced from the
susceptibility and specific heat data, respectively.}
\begin{ruledtabular}
\begin{tabular}{cccccccccc}
sample &$a=b$ &$c$  &$\chi_0$ ($\times10^{-3}$)&$C'$
($\times10^{-3}$) &$\Delta^{1}$(1T) &$\Delta^{2}$(0T)
&$\Delta^{2}$(7T) &$\Delta^{2}$(11T) &$\Delta^{2}$(14T)
\\\cline{2-10}
&(\AA) &(\AA) & (emu/(mol Cu))& (emu K/(mol Cu))&
(K)&(K)&(K)&(K)&(K)  \\\hline
Al& 8.993 & 6.640  & -1.18 & 18.21 & 17.9 & 22.1 & $14.4$ &$10.8$ & $10.0$  \\
La& 8.993 & 6.649 & -1.59 & 14.88& $14.1$ & $19.6$ & $15.6$ & $11.7$ & $8.9$ \\
Na& 8.992 & 6.648  & -1.08 & 15.42 & $19.8$ & $-$& $-$ & $-$ & $-$  \\
Y& 8.992 & 6.642  & -1.30 & 12.45 & $16.4$ & $26.5$ & $20.5$ & $14.9$ & $11.0$  \\
Sr& 8.995 & 6.649  & -0.74 & 10.57 & $21.5$ & $-$ & $-$ &
$-$ & $-$\\
\end{tabular}
\end{ruledtabular}
\end{table*}

The Curie-tail in the magnetic susceptibility is more pronounced
in the Al doped compound than in SrCu$_2$(BO$_3$)$_2$ and other
doping systems. A probable cause for this is that part of
Cu$^{2+}$ within the CuBO$_3$ plane was replaced by Al$^{3+}$ and
the dimerized Cu spin structure was distorted, leading to the
large susceptibility upturn in the low temperature regime.

In order to further understand both the ground states and
low-lying excited states, we measured the specific heat of
M$_{0.1}$Sr$_{0.9}$Cu$_2$(BO$_3$)$_2$ with M=Y, La and Al from
1.9K to 25K in applied fields up to 14T. The specific heat $C$
divided by temperature $T$, $C/T$, is shown in Fig. 5. For all
three samples, a peak in $C/T$ at around 7.5K is observed clearly
in zero field, and the peak is suppressed to lower temperatures in
applied fields. For an example, for
Y$_{0.1}$Sr$_{0.9}$Cu$_2$(BO$_3$)$_2$, the peak positions for
$H$=0T, 7T, 11T, 14T are, 7.4K, 6.8K, 6.2K, 6.1K, respectively.
They are very close to the values for SrCu$_2$(BO$_3$)$_2$
reported previously \cite{Kageyama2, Jorge, Tsujii}. Above 13K,
$C/T$ gradually increases with increasing $T$ since the phonon
contribution becomes the dominant part of total specific heat.
Comparing the magnetic susceptibility $\chi$ and the $C/T$ at zero
field, we find that the peak position of $\chi$ is about twice the
peak temperature of $C/T$. This is due to the fact that $\chi$ is
a measure of two-particle excitations, while $C/T$ is only a
measure of one-particle density of states \cite{Tao}.

\begin{figure}[t]
\includegraphics[width=8cm]{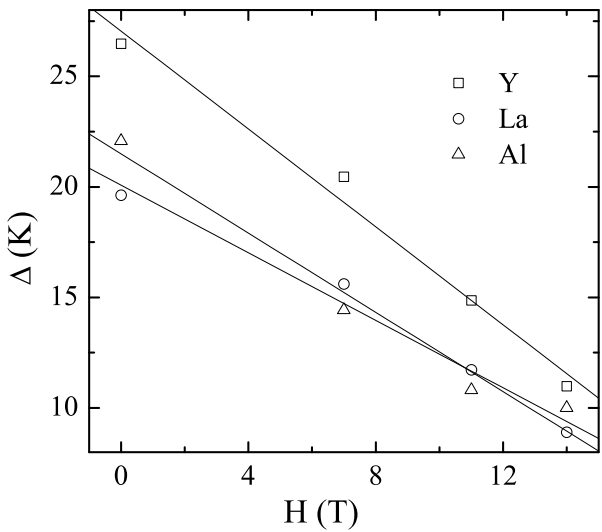}
\caption{\label{fig:liufig_6} {Spin gap $\Delta(H)$ versus field
$H$ for M$_{0.1}$Sr$_{0.9}$Cu$_2$(BO$_3$)$_2$ with M=Y, La and Al.
The solid lines are linear fits to the data points.}}
\end{figure}

In low temperature regime, the specific heat is a sum of the
contribution from low-lying magnetic excitations and the
contribution from phonons. The latter is proportional to $T^3$.
Thus the low temperature specific heat can be expressed as:
\begin{equation}
\emph{C}=\frac{A}{\emph{T}} \exp(-\Delta/\emph{T}) +
\beta\emph{T}^3 .\label{heat}
\end{equation}
By fitting the experimental data with the above equation below 7K,
we obtained the gap value $\Delta$ in different $H$ (Table 1). As
shown in Fig. 6, $\Delta$ drops linearly with $H$. This linear
field dependence of $\Delta$ can be explained by the Zeeman splitting
of excited triplet levels (S=1). From the gap value determined from the
specific heat data, we find that $\Delta$ is suppressed most markedly
in the La-doped sample, in agreement with the susceptibility data.
For the Al-doped sample, the peak value of the specific heat is much
less than those of other samples. It suggests that the Al doping can
suppress most strongly low lying spin excitations in the CuBO$_3$ planes
in comparison with other samples.

Kageyama and co-workers \cite{Kageyama1, Kageyama2} measured the
magnetic susceptibility, Cu NQR, and inelastic neutron scattering
spectra of SrCu$_2$(BO$_3$)$_2$. From the susceptibility data, it
was found that the spin gap is $\sim$19K, in agreement with our
value of the spin gap, $\sim$21K, obtained by the same kind of
measurement. However, the spin gap estimated from the other
measurements is larger than that from the susceptibility. $\Delta$
estimated from the Cu NQR and the inelastic neutron scattering
measurements are about 30K and 34K, respectively. The spin gap
estimated from specific heat \cite{Kageyama3}, ESR \cite{Nojiri}
and Raman scattering \cite{Lemments} measurements agrees with that
from inelastic neutron scattering measurements. Our specific heat
results show that the spin gaps for
M$_{0.1}$Sr$_{0.9}$Cu$_2$(BO$_3$)$_2$ with M=Y, Al and La are
about 27K, 22K and 20K, less than that for SrCu$_2$(BO$_3$)$_2$,
consistent with our susceptibility measurements. However, the spin
gap estimated from the susceptibility measurements is about 10K
less than that from the specific heat measurements.

Recently, electron or hole doping effects on SrCu$_2$(BO$_3$)$_2$
were investigated by a number of authors. Zorko et al.
\cite{Zorko2} investigated the doping effect by immersing
SrCu$_2$(BO$_3$)$_2$ into Li-NH$_3$ solution. They found that the
doping did not change much the magnetic susceptibility behavior,
in agreement with our observation. The substitution of Sr by La or
Ba was tried by Zorko and Ar\v con \cite{Zorko1} and by Choi et
al. \cite{Choi}, respectively. Zorko et al. \cite{Zorko1} also
tried to substitute Cu by Mg. However, no superconductivity was
observed in these doped systems.

\section{\label{sec:level1}Conclusions\protect\\}

In conclusion, SrCu$_2$(BO$_3$)$_2$ and
M$_{0.1}$Sr$_{0.9}$Cu$_2$(BO$_3$)$_2$ with M=Al, La, Na and Y
polycrystalline samples were successfully prepared by solid state
reaction. XRD analysis showed that these doping compounds are
single-phased with similar structure as for SrCu$_2$(BO$_3$)$_2$.
The spin gap is suppressed by doping.
However, there is no superconductivity observed in the four doped samples.
The doping effects on SrCu$_2$(BO$_3$)$_2$ need to be further investigated.

\begin{acknowledgments}
We would like to thank Z. J. Chen and L. Lu for
useful discussions. This work is supported by the National Natural Science
Foundation of China.
\end{acknowledgments}


\begin{thebibliography}{16}

\bibitem{Azuma} M. Azuma, Z. Hiroi, M. Takano, K. Ishida, and Y. Kitaoka, { Phys. Rev. Lett.} {\bf 73}, 3463 (1994).

\bibitem{Iwase} H. Iwase, M. Isobe, Y. Ueda, and H. Yasuoka, { J. Phys. Soc. Jpn.} {\bf 65}, 2397 (1996).

\bibitem{Kageyama1} H. Kageyama, K. Yoshimura, R. Stern, N. V. Mushnikov, K. Onizuka, M. Kato, K. Kosuge, C. P. Slichter, T. Goto, and Y. Ueda, { Phys. Rev. Lett.} {\bf 82}, 3168 (1999).

\bibitem{Smith} R. W. Smith and D. A. Keszler, { J. Solid. State. Chem.} {\bf 93}, 430 (1991).

\bibitem{Miyahara} S. Miyahara and K. Ueda, { Phys. Rev. Lett.} {\bf 82}, 3701 (1999).

\bibitem{Shastry1} B. S. Shastry and B. Sutherland, { Physica (Amsterdam)} {\bf 108B}, 1069 (1981).

\bibitem{Zheng} Zheng. Weihong, C. J. Hammer, and J. Oitmaa, { Phys. Rev. B} {\bf 60}, 6608 (1999).

\bibitem{Shastry2} B. S. Shastry and B. Kumar, { Prog. Theor. Phys. Suppl.} {\bf 145}, 1 (2002).

\bibitem{Kimura} T. Kimura, K. Kuroki, R. Arita, and H. Aoki, { Phys. Rev. B} {\bf 69}, 054501 (2004).

\bibitem{Sorensen} E. S. Sorensen and I. Affleck, { Phys. Rev. Lett.}  {\bf 71}, 1633 (1993).

\bibitem{Tao} T. Xiang, { Phys. Rev. B}  {\bf 58}, 9142 (1998).

\bibitem{Kageyama2} H. Kageyama, M. Nishi, N. Aso, K. Onizuka, T. Yosihama, K. Nukui, K. Kodama, K. Kakurai, and Y. Ueda, { Phys. Rev. Lett.} {\bf 84}, 5876 (2000).

\bibitem{Jorge} G. Jorge, M. Jaime, N. Harrison, R. Stern, H. Dabkowska, and B. D. Gaulin, { J. Alloys. Compd. } {\bf 369}, 90 (2004).

\bibitem{Tsujii} H. Tsujii, R. C. Rotundu, B. Andraka, Y. Takano, H. Kageyama, and Y. Ueda, { Cond-mat/031509}.

\bibitem{Kageyama3} H. Kageyama, H. Suzuki, M. Nohara, K. Onizuka, H. Takagi, and Y. Ueda, { Physica B.} {\bf 281}, 667 (2000).

\bibitem{Nojiri} H. Nojiri, H. Kageyama, K. Onizuka, Y. Ueda, and M. Motokawa, J. Phys. Soc. Jpn. {\bf 68}, 2906 (1999)

\bibitem{Lemments} P. Lemments, M. Grove, M. Fischer, G. Guntherodt, V. N. Kotov, H. Kageyama, K. Onizuka, and Y. Ueda, {Phys. Rev. Lett.} {\bf 85}, 2605 (2000).

\bibitem{Zorko2} A. Zorko, D. Ar\v con, C. J. Nuttall, and A. Lappas, { J. Magn. Magn. Mater.} (in press).

\bibitem{Zorko1} A. Zorko and D. Ar\v con, { Phys. Rev. B}  {\bf 65}, 024417 (2001).

\bibitem{Choi} K. -Y. Choi, Yu. G. Pashkevich, K. V. Lamonova, H. Kageyama, Y. Ueda, and P. Lemmens, { Phys. Rev. B}  {\bf 68}, 104418 (2003).

\end{thebibliography}
\end{document}